\documentclass[letterpaper, 10 pt, conference]{ieeeconf}  

\IEEEoverridecommandlockouts                              

\usepackage{graphicx}
\usepackage{amsmath,amssymb,amsfonts}
\usepackage[caption=false]{subfig}
\usepackage[dvipsnames]{xcolor}

\usepackage{tikz}
\definecolor{darkYellow}{rgb}{0.8,0.65,0}
\definecolor{darkOrchid}{rgb}{0.4,0,1}
\usepackage{multicol}
\usepackage{multirow}
\usepackage{array}

\newcommand{\REV}[1]{\textcolor{black}{#1}}

\newtheorem{problem}{Problem}

\title{\LARGE \bf
Collective Decision Making using Attractive and Repulsive Forces in Markovian Opinion Dynamics*
}

\author{Carl-Johan Heiker$^{1}$ and Paolo Falcone$^{1, \dagger}$
\thanks{*This work is supported by the Swedish Innovation Agency (VINNOVA) under the grant 2018-05005.}
\thanks{$^{1}$Both authors are with the Mechatronics group at Chalmers University of Technology, Gothenburg, Sweden.
        {\tt\small \{heikerc, falcone\}@chalmers.se}}%
\thanks{$^{\dagger}$Paolo Falcone is with the Engineering Department ``Enzo Ferrari'', University of Modena
and Reggio Emilia, Italy.
        {\tt\small falcone@unimore.it}}%
}

\begin{document}


\maketitle
\thispagestyle{empty}
\pagestyle{empty}

\begin{abstract}
In this paper, we model a decision-making process involving a set of interacting agents. We use Markovian opinion dynamics, where each agent switches between decisions according to a continuous time Markov chain.
Existing opinion dynamics models are extended by
introducing attractive and repulsive forces that act within and between groups of agents, respectively. Such an extension enables the resemblance of behaviours emerging in networks where agents make decisions that depend both on their own preferences and the decisions of specific groups of surrounding agents. 
\REV{The considered modeling problem and the contributions in this paper are inspired by the interaction among road users (RUs) at traffic junctions, where each RU has to decide whether to go or to yield.} 
\end{abstract}


\section{INTRODUCTION}
\REV{For autonomous vehicles to be safe, their motion must be planned based on the surrounding human road users' (pedestrians, cyclists, drivers) \emph{future actions}, which are normally unavailable. For this reason, collisions with RUs can only be avoided \emph{in probabilities}. For example, the model predictive planner in \cite{batkovic2021robust} negotiates a crosswalk with an approaching pedestrian, based on \emph{static} probabilities that the pedestrian chooses to cross or stay on the sidewalk. In this paper, our objective is to develop a model which describes the evolution of such probabilities based on the complete traffic scene around the vehicle. Such a model could complement the prediction model used by the planner, in order to plan a path while accounting for the evolution of the surrounding traffic scene. In this paper, we explore the field of \emph{opinion dynamics} to derive that model.} 


Opinion dynamics is a multi-agent modeling framework introduced to explain how opinion spreads in a population.
The overview in \cite{noorazar2020survey} of both classical and modern opinion dynamics describes two main categories of models. First, DeGrootian models \cite{degroot1974consensus} treat the opinion of each agent as a linear combination of its neighbors' opinions. These models have been extended to describe other social phenomena, for example in \cite{friedkin1990social} where agents reluctant to change are considered. Second, in bounded confidence models such as \cite{deffuant2001mixing}, agents only influence each other if their difference in opinions is below some tolerance.

While classic opinion dynamics often assumes that agents select opinions deterministically, processes with agents that decide their opinion state stochastically require different modeling approaches.
One such approach is to represent agents as Markov chains, using either their discrete-time representation (DTMC) as in \cite{asavathiratham2001influence}, or their continuous-time form (CTMC) as in \cite{vanmieghem2011sis}. Additionally, approximation methods are proposed in both \cite{vanmieghem2011sis} and \cite{banisch2012opinion} to address the scalability issues of Markovian networks.
Using the CTMC agent representation, it is possible to describe how agents interact through state transition rate modulation.
Based on a nonlinear approach in  \cite{bolzern2014consensus}, a linear model was introduced in \cite{bolzern2018opinion} and \cite{bolzern2019opinion} to describe opinion configurations as states in a network of Markovian agents. This model was fitted with centralized tuning parameters for influence strength, inter-agent trust and opinion uncertainty in \cite{bolzern2020opinion}, and can in some cases be reduced into a lower order marginalized model. The model has been applied on a political problem in \cite{bolzern2021effect}, and in \cite{farooqi2019railway} to describe the decision process of choosing parameters in a collective energy minimization problem for communicating trains.

The model proposed in this paper builds upon an extension of the framework in \cite{bolzern2018opinion}-\cite{bolzern2020opinion}. 
While the model in~\cite{bolzern2018opinion}-\cite{bolzern2020opinion} 
assumes that agents are only attracted to others' decision states with an influence strength proportional to how many other agents are in that decision state, we introduce a \emph{repulsive action} through rate modulation. A form of repulsion in opinion dynamics was explored in \cite{noorazar2018energy}, while we suggest the addition of a repulsive force specifically adapted to the framework in~\cite{bolzern2018opinion}-\cite{bolzern2020opinion}. 
Furthermore, we propose to divide the set of interacting agents into attractive and repulsive groups, 
as was explored in different ways in \cite{antal2006social}, \cite{altafini2012dynamics} and \cite{yang2019opinion}. Our partition stems from the two interaction types that we consider, where the first describes the attractive forces between agents in each group, while the second expresses the repulsive forces between agents from different groups.

The paper starts by describing Markovian agent networks in Section \ref{sec:network}, before introducing the agent interaction in Section \ref{sec:groups}. We illustrate our model in a traffic intersection example in Section \ref{sec:results}, while the conclusions in Section \ref{sec:conclusions} close the manuscript.

\section{Markovian Agent Networks}
\label{sec:network}

\begin{figure}[t]
    \centering
    \begin{tikzpicture}[thick,scale=0.63, every node/.style={scale=0.8}]
        \draw[line width=0.1mm, solid] (-1, 2.5) -- (-1, 0.5);
        \draw[line width=0.1mm, solid] (-3, 0.5) -- (-1, 0.5);
        \draw[line width=0.1mm, solid] (1, 2.5) -- (1, 0.5);
        \draw[line width=0.1mm, solid] (3, 0.5) -- (1, 0.5);
        \draw[line width=0.1mm, solid] (-3, -1) -- (3, -1);
        \node[rectangle,draw,fill=gray] at (1.5,0) {$\hspace{4pt}3\hspace{4pt}$};
        \node[rectangle,draw,fill=gray] at (2.5,0) {$\hspace{4pt}4\hspace{4pt}$};
        \node[rectangle,draw,fill=gray] at (3.5,0) {$\hspace{4pt}5\hspace{4pt}$};
        \draw [line width=0.5mm] (-2, -0.5) -- (-1, -0.5);
        \draw [line width=0.5mm] (-3.3, -0.5) -- (-2.3, -0.5);
        \node[circle,draw,fill=Goldenrod] at (-1.5,-0.5) {$2$};
        \node[circle,draw,fill=Goldenrod] at (-2.8,-0.5) {$1$};
        \draw [line width=0.5mm] (-0.5, 2.5) -- (-0.5, 1.5);
        \draw [line width=0.5mm] (-0.5, 1.45) -- (-0.5, 0.5);
        \node[circle,draw,fill=Orchid] at (-0.5,1) {$7$};
        \node[circle,draw,fill=Orchid] at (-0.5,2) {$6$};
        \node[] at (0,3.5) {\textit{North}};
        \node[] at (-6,-0.25) {\textit{West}};
        \node[] at (6,-0.25) {\textit{East}};
        \node[] at (0, 4){};
    \end{tikzpicture}
    \caption{
    \REV{Unsignaled T-junction with cyclists (1-2 and 6-7) and drivers (3-4).}}
    \label{fig:intersectionExample}
\end{figure}
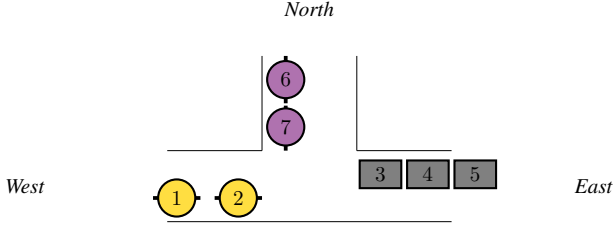

\REV{
In order to clearly illustrate the modeling framework adopted in this paper, we consider the traffic example sketched in Fig. \ref{fig:intersectionExample}, which shows 
an unsignaled T-junction.} Two groups of cyclists are approaching from the west and from the north, respectively, while one group of three drivers is approaching from the east. 
\REV{We view this intersection scenario as a decision process, where 
RUs either \emph{decide to yield or to go} through the intersection} depending on both their destinations and the decisions of other agents. We then state the following problem:
\begin{problem}





\label{problem1}
\REV{A set of RUs approach the intersection at $t_{0}$, at which time their initial decisions are known with some probability. Predict the evolution of the the RUs' most likely decisions at~$t\ge t_0$.}
\end{problem}

To solve \textit{Problem \ref{problem1}}, we start by modeling the decision process of a single agent (RU).

\subsection{Single agent behavior}


An agent $r$ can switch between states in the decision state set 
\begin{equation}
    \mathcal{S} = \{s_{i}\vert i=1,2,\dots, M\},
\end{equation}
at any point in continuous time. 
\REV{In Fig. \ref{fig:intersectionExample}, each agent can choose between the states $s_{1}=\textit{Yield}$ and $s_{2}=\textit{Go}$}.
We describe the stochastic state transitions of $r$ as a CTMC over the set $\mathcal{S}$, with $\pi^{r}_{i}(t)$ being the probability that $r$ is in state $s_{i}$ at time $t$. The state probability vector 
is defined as
\begin{equation}
    \Pi^{r}(t) = \begin{bmatrix}\pi^{r}_{1}(t)&\pi^{r}_{2}(t)&\dots&\pi^{r}_{M}(t)\end{bmatrix}^{T},
\end{equation}
and the CTMC of $r$ is fully defined for a state set $\mathcal{S}$, a state transition rate matrix $Q^{r}\in\mathbb{R}^{M\times M}$ and an initial state probability distribution $\Pi^{r}(0)$ \cite{cassandras2008introduction}. By definition, $Q^{r}$ has positive off-diagonals, and diagonal elements $Q^{r}_{i,i}$ equal to the negated sum of the off-diagonals on row $i$. If the chain is in $s_{i}$, the time until a transition to $s_{j}$ is exponentially distributed with the associated rate parameter $Q^{r}_{i,j}$, so that the probability of two or more transitions occurring at exactly the same time is zero. The time derivative of the state probabilities in a CTMC is
\begin{equation}
    \label{eq:isolatedAgent}
    \dot{\Pi}^{r}(t) = (Q^{r})^{T}\Pi^{r}(t),
\end{equation}
and the state probability trajectories are given by
\begin{equation}
    \label{eq:basicMarkov}
    \Pi^{r}(t) = e^{(Q^{r})^{T}t}\Pi^{r}(0).
\end{equation}
As in \cite{bolzern2018opinion} and related works \cite{bolzern2019opinion}-\cite{bolzern2021effect}, our agent CTMCs are \emph{ergodic}, and therefore irreducible with positive recurrent states. This means in essence that at any time $t$, each CTMC has a nonzero probability to transition to any of its states, and details can be found in \cite{cassandras2008introduction}. This in turn ensures the existence of a unique stationary state probability vector $\bar{\Pi}^{r}$, independent of $\Pi^{r}(0)$, in which all elements are positive. To find it, we may set the dynamics of (\ref{eq:isolatedAgent}) to zero and solve the constrained linear problem
\begin{subequations}
\begin{align}
    &(Q^{r})^{T}\bar{\Pi}^{r} = 0, \label{eq:agentSS1}\\ 
    &\sum_{j=1}^{M}\bar{\pi}^{r}_{j} = 1. \label{eq:agentSS2}
\end{align}
\end{subequations}

\REV{Fig. \ref{fig:single} shows a two-state CTMC for an agent $a$. The transition rates in $Q^{a}$ determine the time until convergence to a stationary state in (\ref{eq:basicMarkov}), and which of the states $s_{1}$ and $s_{2}$ will obtain the highest stationary state probability.} Next, we derive the state transition probabilities of an entire network of Markovian agents.

\REV{
\begin{figure}[t]
\centering
\begin{tikzpicture}[thick,scale=0.8, every node/.style={scale=0.8}]
    \node [draw, circle, fill opacity=1] at (-1, 1) (0.7cm) (S1) {$s_{1}$};
    \node [draw, circle, fill opacity=1] at (1, 1) (0.7cm) (S2) {$s_{2}$};
    \draw (S1) edge[->,bend left, above] node{$Q^{a}_{1,2}$} (S2);
    \draw (S2) edge[->,bend left, below] node{$Q^{a}_{2,1}$} (S1);
    \node[] at (0, 2){};
\end{tikzpicture}
\caption{
\REV{Markov chain for a single agent $a$ with states $\mathcal{S} = \{s_{1}, s_{2}\}$. By convention, self-loops are not drawn.}}
\label{fig:single}
\end{figure}
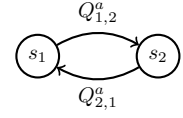
}
\subsection{Network of Markovian agents}
As in \cite{bolzern2018opinion}, we start by considering $N$ Markovian agents. Each agent $r$ is described by an individual CTMC over the shared state space $\mathcal{S}=\{s_{i}\vert i = 1,2,\dots, M\}$. 
However, the transition rates in each $Q^{r}$ are different for each individual $r$.
We formulate a network state as the tuple 
\begin{equation}
    X = \langle s^{1},\dots, s^{r}, \dots,s^{N} \rangle,
\end{equation}
describing a permutation of individual agent states. As each of the $N$ agents can choose among $M$ states, there are $M^N$ network states. When observing transitions in all $N$ CTMCs at the same time, the probability of simultaneous state transitions remains zero due to the exponentially distributed inter-event times of each chain. A transition between two network states is therefore defined by the transition of a single agent. The network is thus also a CTMC with $M^N$ states, fully defined by a set of all network state configurations, rate matrices $Q^{1}, \dots, Q^{N}$ and the individual initial state probabilities. Importantly, the ergodicity of each agent implies that the network CTMC is ergodic too, as described in \cite{bolzern2018opinion}. Like in \cite{bolzern2019opinion}, the time derivative of the network state probabilities is
\begin{equation}
\label{eq:basicIsolatedNetwork}
    \dot{\Pi}_{X}(t) = Q_{0}^{T}\Pi_{X}(t),
\end{equation}
where the ${M^N \times M^N}$ matrix $Q_{0}$ is given as
\begin{equation}
\label{eq:Q0}
    Q_{0} = \sum_{r=1}^{N}I_{M^{r-1}}\otimes Q^{r} \otimes I_{M^{N-r}}.
\end{equation}
\REV{Thus, the evolution of the $M^{N}$ network state probabilities $\Pi_{X}(t)$ and their stationary state values $\bar{\Pi}_{X}$ are found by using $Q_{0}$ in the same equations as for a single agent, (\ref{eq:isolatedAgent}) and (\ref{eq:agentSS1}) with (\ref{eq:agentSS2}), respectively.}

\subsection{Marginalization principle}
\label{sec:margPrinc}
The state space of a CTMC network has dimension $M^N$, which can be large. To alleviate this, marginalized state probabilities are derived in \cite{bolzern2018opinion} and used in the related works \cite{bolzern2019opinion} and \cite{bolzern2020opinion}. The marginalized model is also a linear model, but instead of describing a CTMC over $M^N$ network states, it expresses the $N$ stacked $M\times 1$ state probabilities of each agent, reducing the dimension of the state space to $NM$. The probability that agent $r$ is in state $s_{j}$ in the marginalization can be derived by summing over all network state probabilities regarding states in which $r$ is in $s_{j}$. 
It is of interest to find an analytical expression for the marginalization, which is trivial with isolated agents, as the transition probabilities of one agent are independent of those of other agents. Let us now investigate how to model agents that change their transition probabilities through interaction.


\section{Attraction and Repulsion}


\label{sec:groups}
In \cite{bolzern2018opinion}-\cite{bolzern2020opinion}, agents influence each other through transition rate modulation. On top of an isolated rate from $Q^{r}$, the transition of an agent $r$ gets an additional rate depending on the fraction of neighbors who are currently in the destination state, through an \emph{attractive force function}.
\REV{Importantly, this function only assumes that agents observe others' instant state changes, rather than their state probabilities. In our traffic example, this translates to the ability of road users to intercept and respond to the simple behavioral changes of others, such as braking or accelerating.} Next, we redefine the attractive force for the use of \emph{different agent groups}.

\subsection{Attraction within groups}

A group $\mathcal{A}$ is a subset of $n\leq N$ agents, that influence each other according to the edges $\mathcal{E}^{\mathcal{A}}$ of the graph $\mathcal{G}_{\mathcal{A}} = (\mathcal{A}, \mathcal{E}^{\mathcal{A}}, \Lambda^{\mathcal{A}})$, where $\Lambda^{\mathcal{A}}$ is a weighted, positive and row-normalized adjacency matrix in which each element determines the relative interaction strength between two group members. The attractive force 
\REV{$\psi_{j}^{\mathcal{A}}(r)$} that an agent $r$ in $\mathcal{A}$ experiences towards state $s_{j}$ depending on the other group members $k_{1}$ is expressed as
\REV{
\begin{equation}
    \label{eq:attractiveInfluence}
    \psi_{j}^{\mathcal{A}}(r)= \eta^{r}\lambda^{\mathcal{A}}\sum_{k_{1}\in \mathcal{A}}\Lambda_{r,k_{1}}^{\mathcal{A}}I_{j}^{k_{1}}(t) .
\end{equation}
}The indicator function $I_{j}^{k_{1}}(t)$ is one if a group member $k_{1}$ is in $s_{j}$ at time $t$ and zero if not, and the positive scalar parameter $\eta^{r}$ determines the level of decision uncertainty of agent $r$. The parameter $\lambda^{\mathcal{A}}$, also a positive scalar, describes the influence strength between the members of $\mathcal{A}$.

\begin{figure}[t]
\centering
\begin{tikzpicture}[thick, scale=0.8, every node/.style={scale=0.8}]
    \fill[color=gray, opacity=0.3] (0,0.5) ellipse (1.4cm and 1cm);
    \fill [color=Orchid, opacity=0.3, rotate around={45:(2.45, 0.35)}] (2.45, 0.35) ellipse (0.7cm and 0.9cm);
    \fill [color=Goldenrod, opacity=0.3, rotate around={-45:(-2.45, 0.35)}] (-2.45, 0.35) ellipse (0.7cm and 0.9cm);
    \node [draw, circle, fill=gray, fill opacity=1] at (0, 0) (0.4cm) (4) {$4$};
    \node [draw, circle, fill=gray, fill opacity=1] at (-0.7, 0.7) (0.4cm) (3) {$3$};
    \node [draw, circle, fill=gray, fill opacity=1] at (0.7, 0.7) (0.4) (5) {$5$};
    \node [draw, circle, fill=Orchid, fill opacity=1] at (2.1, 0.7) (0.4cm) (6) {$6$};
    \node [draw, circle, fill=Orchid, fill opacity=1] at (2.8, 0) (0.4cm) (7) {$7$};
    \node [draw, circle, fill=Goldenrod, fill opacity=1] at (-2.1, 0.7) (0.4cm) (2) {$2$};
    \node [draw, circle, fill=Goldenrod, fill opacity=1] at (-2.8, 0) (0.4cm) (1) {$1$};
   \node [] at (-2.7,0.6) (L1){$\mathcal{C}_{1}$};
    \node [] at (2.7,0.6) (L1){$\mathcal{C}_{2}$};
    \node [] at (0,1) (L1){$\mathcal{D}$};
    \draw [line width=0.45mm] (1)--(2);
    \draw [line width=0.45mm] (3)--(4);
    \draw [line width=0.45mm] (3)--(5);
    \draw [line width=0.45mm] (4)--(5);
    \draw [line width=0.45mm] (6)--(7);
    \node [] at (0,1.7) {};
\end{tikzpicture}
\caption{Group $\mathcal{C}_{1}$ contains agents $1$ and $2$, group $\mathcal{D}$ has agents $3$ to $5$ while agents $6$ and $7$ form group $\mathcal{C}_{2}$. Solid lines represent attractive forces.}
\label{fig:attract}
\end{figure}
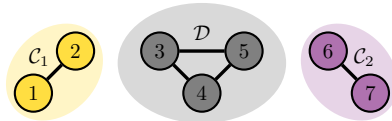

In the intersection example, the cyclists and drivers from Fig. \ref{fig:intersectionExample} are divided into groups by vehicle type and origin. The sets can be seen in Fig. \ref{fig:attract}, where $\mathcal{C}_{1}=\{1, 2\}$ contains the cyclists arriving from the west, $\mathcal{D}= \{3, 4, 5\}$ holds the drivers approaching from the east while $\mathcal{C}_{2}=\{6, 7\}$ describes the cyclists coming from the north. 
\REV{If we for example observe that cyclists in $\mathcal{C}_{1}$ often follow each other through the intersection while drivers in $\mathcal{D}$ do not, we may set $\lambda^{\mathcal{D}} < \lambda^{\mathcal{C}_{1}}$ in the corresponding attractive forces within these groups.}

\subsection{Repulsion between groups}



\begin{figure}[b]
\centering
\begin{tikzpicture}[thick, scale=0.8, every node/.style={scale=0.8}]
    \fill[color=gray, opacity=0.3] (0,0.5) ellipse (1.4cm and 1cm);
    \fill [color=Orchid, opacity=0.3, rotate around={45:(2.45, 0.35)}] (2.45, 0.35) ellipse (0.7cm and 0.9cm);
    \fill [color=Goldenrod, opacity=0.3, rotate around={-45:(-2.45, 0.35)}] (-2.45, 0.35) ellipse (0.7cm and 0.9cm);
    \node [draw, circle, fill=gray, fill opacity=1] at (0, 0) (0.4cm) (4) {$4$};
    \node [draw, circle, fill=gray, fill opacity=1] at (-0.7, 0.7) (0.4cm) (3) {$3$};
    \node [draw, circle, fill=gray, fill opacity=1] at (0.7, 0.7) (0.4) (5) {$5$};
    \node [draw, circle, fill=Orchid, fill opacity=1] at (2.1, 0.7) (0.4cm) (6) {$6$};
    \node [draw, circle, fill=Orchid, fill opacity=1] at (2.8, 0) (0.4cm) (7) {$7$};
    \node [draw, circle, fill=Goldenrod, fill opacity=1] at (-2.1, 0.7) (0.4cm) (2) {$2$};
    \node [draw, circle, fill=Goldenrod, fill opacity=1] at (-2.8, 0) (0.4cm) (1) {$1$};
    \node [] at (-2.7,0.6) (L1){$\mathcal{C}_{1}$};
    \node [] at (2.7,0.6) (L1){$\mathcal{C}_{2}$};
    \node [] at (0,1) (L1){$\mathcal{D}$};
    \draw[line width=0.45mm,dashed, color=darkOrchid] (3) to [out=45,in=90+20] (6);
    \draw[line width=0.45mm,dashed, color=darkOrchid] (3) to [out=-17,in=190] (7);
    \draw[line width=0.45mm,dashed, color=darkOrchid] (4) to [out=90,in=90+35](6);
    \draw[line width=0.45mm,dashed, color=darkOrchid] (4) to [out=0,in=210] (7);
    \draw[line width=0.45mm,dashed, color=darkOrchid] (5) to [out=45,in=90+55] (6);
    \draw[line width=0.45mm,dashed, color=darkOrchid] (5) to [out=-35,in=170] (7);
    \draw[line width=0.45mm,dashed, color=darkYellow] (3) to [out=135,in=35] (2);
    \draw[line width=0.45mm,dashed, color=darkYellow] (3) to [out=215,in=10] (1);
    \draw[line width=0.45mm,dashed, color=darkYellow] (4) to [out=90,in=55](2);
    \draw[line width=0.45mm,dashed, color=darkYellow] (4) to [out=180,in=-30] (1);
    \draw[line width=0.45mm,dashed, color=darkYellow] (5) to [out=135,in=70] (2);
    \draw[line width=0.45mm,dashed, color=darkYellow] (5) to [out=197,in=-10] (1);
\end{tikzpicture}
\caption{Agents in group $\mathcal{D}$ are repulsed by agents from groups $\mathcal{C}_{1}$ and $\mathcal{C}_{2}$, and vice versa.}
\label{fig:repulsive}
\end{figure}
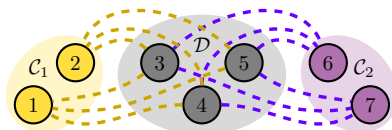

Agents in a group $\mathcal{A}$ can also experience a repulsive form of transition rate influence through a \emph{repulsive force function}, from other groups $\mathcal{R}_{\ell}$ in a set of repulsive groups $R$. We describe this interaction using the graph $\mathcal{G}_{\mathcal{A} \mathcal{R}_{\ell}} = (\mathcal{A}\cup \mathcal{R}_{\ell} , \mathcal{E}^{\mathcal{A} \mathcal{R}_{\ell}}, \Gamma^{\mathcal{A} \mathcal{R}_{\ell}})$, where agents from groups~$\mathcal{A}$ and~$\mathcal{R}_{\ell}$ interact according to the edges in $\mathcal{E}^{\mathcal{A} \mathcal{R}_{\ell}}$ with relative interaction strengths from the positive, weighted, row normalized adjacency matrix $\Gamma^{\mathcal{A} \mathcal{R}_{\ell}}$.
The repulsive force that pushes agent $r$ away from state $s_{i}$ is a function of the number of agents in $\mathcal{R}_{\ell}$ whose state is $s_{i}$. This is expressed as a reactive force 
\REV{
\begin{equation}
    \label{eq:repulsiveInfluence}
    \xi_{j}^{\mathcal{A}\mathcal{R_{\ell}}}(r)= \frac{\eta^{r}\gamma^{\mathcal{A}\mathcal{R}_{\ell}}}{|R|}\sum_{k_{2}\in \mathcal{R}_{\ell}}\Gamma_{r,k_{2}}^{\mathcal{A}\mathcal{R}_{\ell}}\big(1 - I^{k_{2}}_{j}(t)\big),
\end{equation}
}towards state $s_{j}$, where the positive scalar parameter $\gamma^{\mathcal{A}\mathcal{R}_{\ell}}$ describes the repulsion strength from agents in~$\mathcal{R}_{\ell}$ to agents in~$\mathcal{A}$, and the number of repulsive groups $\vert R\vert$ is used for normalization.
\REV{By not expressing the repulsive force as a rate decrease to $s_{i}$, we avoid accidentally introducing negative transition rates to a state. Through (\ref{eq:repulsiveInfluence}), we can for example express that drivers avoid entering the intersection at the same time as cyclists to reduce the risk of collision.} This case is visualized in Fig. \ref{fig:repulsive}, where dashed lines represent that there are repulsive forces between $\mathcal{D}$ and $\mathcal{C}_{1,2}$, but not between $\mathcal{C}_{1}$ and $\mathcal{C}_{2}$. As in the attractive force function, the parameters of the repulsive force function can be set high or low to best describe an observed behavior.

\subsection{Attraction and repulsion in the network model}
The forces (\ref{eq:attractiveInfluence}) and (\ref{eq:repulsiveInfluence}) can be evaluated for every network transition by considering the source and destination states. We can then construct two additional transition rate matrices $A_{0}$ and $R_{0}$ describing attractive and repulsive forces between network states, respectively. 
\REV{Specifically, if a transition between network states $X_{i}$ and $X_{j}$ is caused by agent $r$ transitioning to state $s_{a}$, then $A_{0}(i,j) = \psi_{a}^{\mathcal{A}}(r)$ while $R_{0}(i,j)=\sum_{\mathcal{R}_{\ell}\in R}\xi_{a}^{\mathcal{A}\mathcal{R_{\ell}}}(r)$.} The network CTMC in (\ref{eq:basicIsolatedNetwork}) can then be reformulated as
\begin{equation}
\label{eq:networkStateEquation}
    \dot{\Pi}_{X}(t) = \big(Q_{0} + A_{0} + R_{0}\big)^{T}\Pi_{X}(t).
\end{equation}
Because $A_{0}$ and $R_{0}$ are both Metzler matrices that merely add rates to the same transitions introduced by $Q_{0}$, (\ref{eq:networkStateEquation}) describes an ergodic CTMC. To translate network state probabilities produced by this system to the marginalized state probabilities according to the summation described in Section \ref{sec:margPrinc}, a matrix $S_{X:m}$ can be constructed such that
\begin{equation}
\label{eq:translation}
    \Pi_{m}(t) = S_{X:m} \Pi_{X}(t).
\end{equation}

\subsection{Marginalization of the network intent model}
The marginalized model describes the time derivative of the probability that each agent $r$ is in state $s_{j}$ at time $t$ when exposed to attractive and repulsive forces $\psi_{j}^{\mathcal{A}}(r)$ and $\xi_{j}^{\mathcal{A}\mathcal{R_{\ell}}}(r)$ as
\begin{equation}
\label{eq:finalExpression}
    \begin{split}
         \dot{\pi}^{r}_{j} &= \sum^{M}_{i=1} Q_{i,j}^{r}\pi_{i}^{r} + \lambda^{\mathcal{A}}\eta^{r} \Big [ \sum_{k_{1} \in \mathcal{A} }\Lambda^{\mathcal{A}}_{r,k_{1}} \pi_{j}^{k_{1}}- \pi_{j}^{r}\Big] +\\
         &-\frac{\eta^{r}}{\vert R \vert} \sum_{\mathcal{\mathcal{R}_{\ell}}\in R}\gamma^{\mathcal{A}\mathcal{R}_{\ell}} \Big[ \sum_{k_{2} \in \mathcal{R}_{\ell}}\Gamma^{\mathcal{A}\mathcal{R}_{\ell}}_{r,k_{2}}\pi_{j}^{k_{2}} + (M-1)\pi_{j}^{r}\Big]\\
        &+ \frac{\eta^{r}}{\vert R \vert} \sum_{\mathcal{\mathcal{R}_{\ell}}\in R}\gamma^{\mathcal{A}\mathcal{R}_{\ell}},
    \end{split}
\end{equation}
where $\pi(t)$ is exchanged for $\pi$ to save space. 
Next, we derive (\ref{eq:finalExpression}).

We use the expected value of indicator functions to denote a state probability, and express the probability that agent $r$ is in state $s_{j}$ in after $\delta t$ time using the infinitesimal defintion of the CTMC. We have that
\begin{equation}
\begin{split}
    \mathbb{E}\big[ I_{j}^{r}(t+\delta t)\vert X(t)\big] &= I_{j}^{r}(t)\big( 1-\delta t \mathbb{Q}_{O}\big)\\ &+ \big( 1-I_{j}^{r}(t) \big ) \delta t \mathbb{Q}_{I},
    \end{split}
\end{equation}
where we denote the sum of outgoing transition rates from $s_{j}$ as $\mathbb{Q}_{O}$ and the sum of incoming rates from other states $s_{i}$ as $\mathbb{Q}_{I}$. By taking the expected value of the RHS, negating $\mathbb{E}\big[ I_{j}^{r}(t)\big]$ and dividing by $\delta t$, we obtain
\begin{equation}
    \label{proof:1}
    \begin{split}
    &\frac{ \mathbb{E}\big[ I_{j}^{r}(t+\delta t)\vert X(t)\big] - \mathbb{E}\big[ I_{j}^{r}(t)\big]}{\delta t}= \\
    &=\mathbb{E}\big[ -I_{j}^{r}(t) \mathbb{Q}_{O}+ \big ( 1-I_{j}^{r}(t)\big ) \mathbb{Q}_{I}\big]= \\
    &=\mathbb{E}\big [ -I_{j}^{r}(t)\big( \mathbb{Q}_{O} + \mathbb{Q}_{I}\big) + \mathbb{Q}_{I} \big ].
    \end{split}
\end{equation}
We note that if we use $\mathbb{E}\big [ I_{j}^{r}(t)\big ]=\pi_{j}^{r}(t)$ and let $\delta t$ approach zero, we obtain an expression for the time derivative $\dot{\pi}_{j}^{r}(t)$ through (\ref{proof:1}). Next, we split the incoming and outgoing rates $\mathbb{Q}_{I}$ and $\mathbb{Q}_{O}$ into three parts. The first part consists of individual transition rates, while the second and third part consist of additive transition rates from the attractive and repulsive forces, respectively. Next, we derive the contributions to the state probability derivative $\dot{\pi}_{j}^{r}(t)$ for each part.

First, we set $\mathbb{Q}_{O} = \sum_{i\neq j}Q_{j,i}^{r}$ and $\mathbb{Q}_{I}=\sum_{i\neq j}Q_{i,j}^{r}I_{i}^{r}(t)$ in the RHS of \ref{proof:1}, which becomes
\begin{equation}
    \begin{split}
        \mathbb{E}\Big [ -I_{j}^{r}(t)\Big ( \sum_{i\neq j}Q_{j,i}^{r} + \sum_{i \neq j} Q_{i,j}^{r}I_{i}^{r}(t)\Big ) + \sum_{i\neq j}Q_{i,j}^{r}I_{i}^{r}(t)\Big ].
    \end{split}
\end{equation}
Since $\sum_{i\neq j}Q_{j,i}^{r} = -Q_{j,j}^{r}$, and $I_{j}^{r}(t)I_{i}^{r}(t) = 0$ for all $t$, the previous expression is reduced to
\begin{equation}
    \begin{split}
        \mathbb{E}\Big [ Q_{j,j}^{r}I_{j}^{r}(t) + \sum_{i\neq j}Q_{i,j}^{r}I_{i}^{r}(t)\Big ] = \mathbb{E}\Big [ \sum_{i=1}^{M}Q_{i,j}^{r}I_{i}^{r}(t)\Big ].
    \end{split}
\end{equation}
Thus, using $\mathbb{E}\big[ I_{i}^{r}(t)\big]=\pi_{i}^{r}(t)$, the contribution to $\dot{\pi}_{j}^{r}(t)$ from the isolated rates is
\begin{equation}
    \sum_{i=1}^{M}Q^{r}_{i,j}\pi_{i}^{r}(t),
\end{equation}
which corresponds to the first term in (\ref{eq:finalExpression}).

Second, we evaluate the rate contributions from the attractive force (\ref{eq:attractiveInfluence}) and set $\mathbb{Q}_{O} = \sum_{i\neq j}\psi_{i}^{\mathcal{A}}(r)$ and $\mathbb{Q}_{I} = \psi_{j}^{\mathcal{A}}(r)$. The RHS of (\ref{proof:1}) becomes
\begin{equation}
\label{proof:2}
    \begin{split}
        &\mathbb{E}\Big [-I_{j}^{r}(t)\Big( \sum_{i\neq j}\psi_{i}^{\mathcal{A}}(r) + \psi_{j}^{\mathcal{A}}(r)\Big) + \psi_{j}^{\mathcal{A}}(r) \Big]= \\
        &=  \mathbb{E}\Big [ -I_{j}^{r}(t)\sum_{i=1}^{M}\psi_{i}^{\mathcal{A}}(r)+\psi_{j}^{\mathcal{A}}(r) \Big].
    \end{split}
\end{equation}
We have that
\begin{equation}
\label{proof:3}
    \begin{split}
        \sum_{i=1}^{M}\psi_{i}^{\mathcal{A}}(r) &= \sum_{i=1}^{M}\lambda^{\mathcal{A}}\eta^{r}\sum_{k_{1}\in \mathcal{A}}\Lambda_{r,k_{1}}^{\mathcal{A}}I_{i}^{k_{1}}(t) =\\
        &=\lambda^{\mathcal{A}}\eta^{r}\sum_{k_{1}\in \mathcal{A}}\Lambda_{r,k_{1}}^{\mathcal{A}}\sum_{i=1}^{M}I_{i}^{k_{1}}(t),
    \end{split}
\end{equation}
where $I_{i}^{k_{1}}(t)=1$ for one and only one $i$ at the same time. Thus, $\sum_{i=1}^{M}I_{i}^{k_{1}}(t)=1$ for any $i$. This holds for every agent $k_{1}\in \mathcal{A}$, so that (\ref{proof:3}) reduces to
\begin{equation}
    \sum_{i=1}^{M}\psi_{i}^{\mathcal{A}}(r) = \lambda^{\mathcal{A}}\eta^{r}\sum_{k_{1}\in \mathcal{A}}\Lambda_{r,k_{1}}^{\mathcal{A}} = \lambda^{\mathcal{A}}\eta^{r},
\end{equation}
because $\Lambda_{r,k_{1}}^{\mathcal{A}}$ is row normalized. Inserting this and writing out $\psi_{j}^{\mathcal{A}}(r)$ in (\ref{proof:2}) yields
\begin{equation}
    \mathbb{E}\Big [-I_{j}^{r}(t)\lambda^{\mathcal{A}}\eta^{r} + \lambda^{\mathcal{A}}\eta^{r}\sum_{k_{1}\in\mathcal{A}}\Lambda_{r,k_{1}}^{\mathcal{A}}I_{j}^{k_{1}}(t)\Big ].
\end{equation}
Using $\mathbb{E}\big[ I_{j}^{r}(t)\big]=\pi_{j}^{r}(t)$, the contribution to $\dot{\pi}_{j}^{r}(t)$ from the attractive force is
\begin{equation}
    \lambda^{\mathcal{A}}\eta^{r}\Big [ \sum_{k_{1}\in \mathcal{A}}\Lambda_{r,k_{1}}^{\mathcal{A}}\pi_{j}^{k_{1}}(t) - \pi_{j}^{r}(t)\Big ],
\end{equation}
which corresponds to the second term in (\ref{eq:finalExpression}).

Lastly, we evaluate the rate contributions from the repulsive force (\ref{eq:repulsiveInfluence}) and set $\mathbb{Q}_{O}=\sum_{\mathcal{R}_{\ell}\in R}\sum_{i\neq j}\xi_{i}^{\mathcal{A}\mathcal{R}_{\ell}}(r)$ and $\mathbb{Q}_{I}=\sum_{\mathcal{R}_{\ell}\in R}\xi_{j}^{\mathcal{A}\mathcal{R}_{\ell}}(r)$. From this, we obtain the RHS of (\ref{proof:1}) as
\begin{equation}
    \begin{split}
        \mathbb{E}\Big[ -I_{j}^{r}(t)\Big( \sum_{\mathcal{R}_{\ell}\in R}\sum_{i\neq j}\xi_{i}^{\mathcal{A}\mathcal{R}_{\ell}}(r) + \sum_{\mathcal{R}_{\ell}\in R}\xi_{j}^{\mathcal{A}\mathcal{R}_{\ell}}(r)\Big) +& \\ +\sum_{\mathcal{R}_{\ell}\in R}\xi_{j}^{\mathcal{A}\mathcal{R}_{\ell}}(r)\Big], \\
    \end{split}
\end{equation}
which may be reduced to
\begin{equation}
\label{proof:4}
    \mathbb{E}\Big[ -I_{j}^{r}(t)\Big( \sum_{\mathcal{R}_{\ell}\in R}\sum_{i=1}^{M}\xi_{i}^{\mathcal{A}\mathcal{R}_{\ell}}(r)  \Big)+\sum_{\mathcal{R}_{\ell}\in R}\xi_{j}^{\mathcal{A}\mathcal{R}_{\ell}}(r)\Big].
\end{equation}
We have that
\begin{equation}
\label{proof:5}
\begin{split}
    \sum_{i=1}^{M}\xi_{i}^{\mathcal{A}\mathcal{R}_{\ell}}(r) = \sum_{i=1}^{M}\frac{\eta^{r}\gamma^{\mathcal{A}\mathcal{R}_{\ell}}}{\vert R \vert}\sum_{k_{2}\in \mathcal{R}_{\ell}}\Gamma_{r,k_{2}}^{\mathcal{A}\mathcal{R}_{\ell}}\Big(1 - I^{k_{2}}_{i}(t)\Big).
\end{split}
\end{equation}
The RHS of this expression is equivalent to
\begin{equation}
\begin{split}
    &\frac{\eta^{r}\gamma^{\mathcal{A}\mathcal{R}_{\ell}}}{\vert R \vert}\sum_{k_{2}\in \mathcal{R}_{\ell}}\Gamma_{r,k_{2}}^{\mathcal{A}\mathcal{R}_{\ell}}\sum_{i=1}^{M}\big(1 - I^{k_{2}}_{i}(t)\big),
\end{split}
\end{equation}
and as $I^{k_{2}}_{i}(t)=1$ for one and only one $i$, it can be simplified to
\begin{equation}
\begin{split}
    &\frac{\eta^{r}\gamma^{\mathcal{A}\mathcal{R}_{\ell}}}{\vert R \vert}\sum_{k_{2}\in \mathcal{R}_{\ell}}\Gamma_{r,k_{2}}^{\mathcal{A}\mathcal{R}_{\ell}}(M-1).
\end{split}
\end{equation}
As $\sum_{k_{2}\in \mathcal{R}_{\ell}}\Gamma_{r,k_{2}}^{\mathcal{A}\mathcal{R}_{\ell}}=1$ due to $\Gamma_{r,k_{2}}^{\mathcal{A}\mathcal{R}_{\ell}}$ being a row normalized matrix, (\ref{proof:5}) becomes
\begin{equation}
\begin{split}
    \sum_{i=1}^{M}\xi_{i}^{\mathcal{A}\mathcal{R}_{\ell}}(r) = \frac{\eta^{r}\gamma^{\mathcal{A}\mathcal{R}_{\ell}}}{\vert R \vert}(M-1).
\end{split}
\end{equation}
We insert this in (\ref{proof:4}) and obtain
\begin{equation}
    \begin{split}
        &\mathbb{E}\Big [ -I_{j}^{r}(t)\Big ( \sum_{\mathcal{R}_{\ell}\in R}\frac{\eta^{r}\gamma^{\mathcal{A}\mathcal{R}_{\ell}}}{\vert R \vert}(M-1)\Big ) + \\
        &+ \sum_{\mathcal{R}_{\ell}\in R} \frac{\eta^{r}\gamma^{\mathcal{A}\mathcal{R}_{\ell}}}{|R|}\sum_{k_{2}\in \mathcal{R}_{\ell}}\Gamma_{r,k_{2}}^{\mathcal{A}\mathcal{R}_{\ell}}\big(1 - I^{k_{2}}_{j}(t)\big)
        \Big ].
    \end{split}
\end{equation}
We move the summation over groups $\mathcal{R}_{\ell}$ and obtain
\begin{equation}
    \begin{split}
        &\sum_{\mathcal{R}_{\ell}\in R}\frac{\eta^{r}\gamma^{\mathcal{A}\mathcal{R}_{\ell}}}{\vert R \vert}\mathbb{E}\Big [ -I_{j}^{r}(t) (M-1) + \\
        &+ \sum_{k_{2}\in \mathcal{R}_{\ell}}\Gamma_{r,k_{2}}^{\mathcal{A}\mathcal{R}_{\ell}}\big(1 - I^{k_{2}}_{j}(t)\big)
        \Big ],
    \end{split}
\end{equation}
which is equivalently expressed as
\begin{equation}
    \begin{split}
        &-\frac{\eta^{r}}{\vert R \vert}\sum_{\mathcal{R}_{\ell}\in R}\gamma^{\mathcal{A}\mathcal{R}_{\ell}} \mathbb{E}\Big[ \sum_{k_{2}\in \mathcal{R}_{\ell}}\Gamma_{r,k_{2}}^{\mathcal{A}\mathcal{R}_{\ell}}I_{j}^{k_{2}}(t) + (M-1)I_{j}^{r}(t)\Big] + \\
        &+ \frac{\eta^{r}}{\vert R \vert}\sum_{\mathcal{R}_{\ell}\in R} \gamma^{\mathcal{A}\mathcal{R}_{\ell}} \sum_{k_{2}\in \mathcal{R}_{\ell}}\Gamma_{r,k_{2}}^{\mathcal{A}\mathcal{R}_{\ell}}.
    \end{split}
\end{equation}
Using the fact that $\Gamma_{r,k_{2}}^{\mathcal{A}\mathcal{R}_{\ell}}$ is row normalized and that $\mathbb{E}\big[ I_{j}^{k_{2}}(t)\big] = \pi_{j}^{k_{2}}(t)$ and $\mathbb{E}\big[ I_{j}^{r}(t)\big] = \pi_{j}^{r}(t)$, we obtain
\begin{equation}
    \begin{split}
        &-\frac{\eta^{r}}{\vert R \vert}\sum_{\mathcal{R}_{\ell}\in R}\gamma^{\mathcal{A}\mathcal{R}_{\ell}} \Big [\sum_{k_{2}\in \mathcal{R}_{\ell}}\Gamma_{r,k_{2}}^{\mathcal{A}\mathcal{R}_{\ell}}\pi_{j}^{k_{2}}(t) + (M-1)\pi_{j}^{r}(t)\Big ] + \\
        &+ \frac{\eta^{r}}{\vert R \vert}\sum_{\mathcal{R}_{\ell}\in R} \gamma^{\mathcal{A}\mathcal{R}_{\ell}}.
    \end{split}
\end{equation}
These terms correspond to the two final terms in (\ref{eq:finalExpression}), which concludes the derivation.

The entire marginalized model can be written in vector form as
\begin{equation}
\label{eq:margModelDE}
    \begin{split}
    \dot{\Pi}_{m}(t) &= (Q_{m} + A_{m} + R_{m})\Pi_{m}(t) + E_{m},
    \end{split}
\end{equation}
where $Q_{m}$, $A_{m}$ and $R_{m}$ are $NM \times NM$ matrices used to collect the internal rates and the terms from the attractive and repulsive forces in (\ref{eq:finalExpression}) for each agent $r$ and state $s_{j}$. $E_{m}$ has dimension $NM\times 1$ and collects the constant terms. Importantly, (\ref{eq:finalExpression}) is an analytical expression that is equivalent to summing over all elements of $\Pi_{X}(t)$ in (\ref{eq:networkStateEquation}) that concern network states in which $r$ is in $s_{j}$. Thus, if this is solved from the initial state $\Pi_{m}(0) = S_{X:m}\Pi_{X}(0)$, the trajectories in (\ref{eq:translation}) are obtained. Compared to the network model, the marginalized model has a much lower state space dimension, but can not distinguish between individual network configurations. Although this generally constitutes a trade-off between low computational complexity and resolution, the marginalized model is preferable for applications that only concerns state probabilities of individual agents.

In \cite{bolzern2019opinion}, the existence of a unique stationary state marginalized probability vector $\bar{\Pi}_{m}$ follows from the fact that the network CTMC is ergodic and has unique stationary network solution. Because our extended network model (\ref{eq:networkStateEquation}) maintains the ergodicity property, the marginalization (\ref{eq:margModelDE}) also has a unique stationary state solution, found by solving
\begin{subequations}
    \begin{align}
        &(Q_{m} + A_{m} + R_{m})\bar{\Pi}_{m} + E_{m} = 0, \label{eq:margModelSS1}\\
        &\sum_{j=1}^{M}\bar{\pi}^{r}_{j} = 1, \quad \forall r. \label{eq:margModelSS2}
    \end{align}
\end{subequations}
\section{Results}
\label{sec:results}

\REV{In this section, we use the traffic intersection problem to illustrate the effects that the rate influence functions (\ref{eq:attractiveInfluence}) and (\ref{eq:repulsiveInfluence}) have on the road users' decision probabilities. This is done by comparing the stationary probabilities of the isolated RUs against the probabilities of the same RUs in the network representation of the traffic scene.
}

\subsection{Problem setup}
\label{sec:resultSetup}
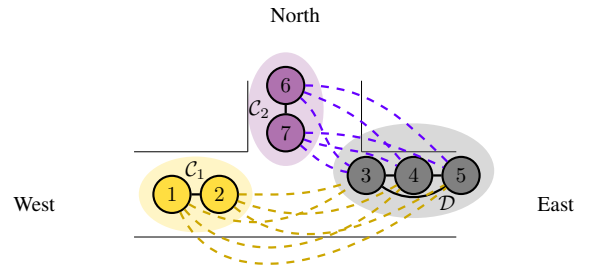
\begin{figure}[b]
    \centering
    \begin{tikzpicture}[thick,scale=0.63, every node/.style={scale=0.8}]
        \draw[line width=0.1mm, solid] (-1, 2.6) -- (-1, 1.1);
        \draw[line width=0.1mm, solid] (-3.4, 1.1) -- (-1, 1.1);
        \draw[line width=0.1mm, solid] (1.4, 2.6) -- (1.4, 1.1);
        \draw[line width=0.1mm, solid] (3.4, 1.1) -- (1.4, 1.1);
        \draw[line width=0.1mm, solid] (-3.4, -0.7) -- (3.4, -0.7);
        \fill[color=Orchid, opacity=0.3] (-0.2, 2) ellipse (0.8cm and 1.2cm);
        \fill [color=gray, opacity=0.3] (2.5, 0.7) ellipse (1.7cm and 1cm);
        \fill [color=Goldenrod, opacity=0.3] (-2.1, 0.2) ellipse (1.2cm and 0.8cm);
        \node[circle,draw,fill=gray] at (1.5,0.6) (3) {$3$};
        \node[circle,draw,fill=gray] at (2.5,0.6) (4) {$4$};
        \node[circle,draw,fill=gray] at (3.5,0.6) (5) {$5$};
        \node[circle,draw,fill=Goldenrod] at (-1.6,0.2) (2) {$2$};
        \node[circle,draw,fill=Goldenrod] at (-2.6,0.2) (1) {$1$};
        \node[circle,draw,fill=Orchid] at (-0.2,1.5)(7) {$7$};
        \node[circle,draw,fill=Orchid] at (-0.2,2.5)(6) {$6$};
        \node[] at (0,4) {$\text{North}$};
        \node[] at (-5.5,0) {$\text{West}$};
        \node[] at (5.5,0) {$\text{East}$};
        \node [] at (-2.1,0.7) (L1){$\mathcal{C}_{1}$};
        \node [] at (-0.75,2) (L2){$\mathcal{C}_{2}$};
        \node [] at (3.2,0.03) (L3){$\mathcal{D}$};
        \draw[line width=0.3mm, solid] (1) to [out=0,in=180] (2);
        \draw[line width=0.3mm, solid] (6) to [out=-90,in=90] (7);
        \draw[line width=0.3mm, solid] (3) to [out=0,in=180] (4);
        \draw[line width=0.3mm, solid] (4) to [out=0,in=180] (5);
        \draw[line width=0.3mm, solid] (5) to [out=-145,in=-35] (3);
        \draw[line width=0.3mm, dashed, color=darkYellow] (1) to [out=-30,in=215] (3);
        \draw[line width=0.3mm, dashed, color=darkYellow] (1) to [out=-50,in=225] (4);
        \draw[line width=0.3mm, dashed, color=darkYellow] (1) to [out=-65,in=220] (5);
        \draw[line width=0.3mm, dashed, color=darkYellow] (2) to [out=0,in=200] (3);
        \draw[line width=0.3mm, dashed, color=darkYellow] (2) to [out=-20,in=210](4);
        \draw[line width=0.3mm, dashed, color=darkYellow] (2) to [out=-40,in=210] (5);
        \draw[line width=0.3mm, dashed, color=darkOrchid] (6) to [out=-40,in=150](3);
        \draw[line width=0.3mm, dashed, color=darkOrchid] (6) to [out=-20,in=130] (4);
        \draw[line width=0.3mm, dashed, color=darkOrchid] (6) to [out=0,in=135] (5);
        \draw[line width=0.3mm, dashed, color=darkOrchid] (7) to [out=-40,in=170] (3);
        \draw[line width=0.3mm, dashed, color=darkOrchid] (7) to [out=-20,in=150] (4);
        \draw[line width=0.3mm, dashed, color=darkOrchid] (7) to [out=0,in=155] (5);
    \end{tikzpicture}
    \caption{Intersection scene with cyclists as groups $\mathcal{C}_{1}$ and $\mathcal{C}_{2}$ and drivers as group $\mathcal{D}$.}
    \label{fig:intersectionExampleResults}
\end{figure}

\REV{In Fig. \ref{fig:intersectionExampleResults}, the road users in the intersection problem are shown. For each RU $r$, the transition rate matrix $Q^{r}$ is defined such that $r$ has a preferred state in the common state space $\mathcal{S} = \{s_{1}, s_{2}\}$, where we recall that $s_{1}=\textit{Yield}$ and $s_{2}=\textit{Go}$}. In addition, drivers are very uncertain while cyclists are more confident. This behavior is achieved by setting $\eta^{r} = 100$ for $r=3,4,5$, $\eta^{r}=10$ for $r=1,6$, and $\eta^{r}=1$ for $r=2,7$.
\REV{Moreover, both drivers and cyclists prefer going instead of yielding, with the exception of $4$, who is a very careful driver.}

Furthermore, Fig. \ref{fig:intersectionExampleResults} shows how attractive and repulsive forces affect each RU in the traffic scene. Attraction occurs within all groups, while the drivers in group $\mathcal{D}$ experience a mutual repulsion from groups $\mathcal{C}_{1}$ and $\mathcal{C}_{2}$. Moreover, the attraction strength $\lambda$ is relatively high within cyclist groups $\mathcal{C}_{1}$ and $\mathcal{C}_{2}$, and low in the driver group $\mathcal{D}$. In addition, the drivers in $\mathcal{D}$ experience a much higher repulsion strength from groups $\mathcal{C}_{1}$ and $\mathcal{C}_{2}$ than vice versa. \REV{Finally, the parameters $\lambda$ and $\gamma$ are chosen such that the transients of the isolated and the non-isolated models show a similar convergence rate, thus resulting in an easy comparison.} Uncertainty parameter values and preferred states are collected in Table \ref{tab:agentParams}, while the attractive and repulsive influence strength parameters can be found in Table \ref{tab:groupParams}.

\begin{table}[b]
\centering
\caption{
Road user parameters}
\label{tab:agentParams}
\begin{tabular}{ p{1.3cm}| p{0.5cm} p{0.5cm} p{0.5cm} p{0.5cm} p{0.5cm} p{0.5cm} p{0.5cm} }
 \multicolumn{8}{c}{Road user} \\
  & 1 & 2 & 3 & 4 & 5 & 6 & 7\\
  \hline
  Pref. state & \REV{$s_{2}$} & \REV{$s_{2}$} & \REV{$s_{2}$} & \REV{$s_{1}$} & \REV{$s_{2}$} & \REV{$s_{2}$} & \REV{$s_{2}$} \\
  $\eta^{r}$ & $10$ & $1$ & $100$ & $100$ & $100$ & $10$ & $1$ \\
\end{tabular}
\end{table}
\begin{table}[b]
\centering
\caption{Group parameters}
\label{tab:groupParams}
\begin{tabular}{p{0.3cm}| p{2cm} p{2cm} p{2cm}}
  \multicolumn{4}{c}{Group} \\
  & $\mathcal{C}_{1}$ & $\mathcal{D}$ & $\mathcal{C}_{2}$ \\
  \hline 
  $\lambda$ & \REV{$0.5$} & \REV{$0.05$} & \REV{$0.5$} \\
  $\gamma$ & \REV{$\gamma^{\mathcal{C}_{1}\mathcal{D}}=0.003$} &
  \REV{$ \gamma^{\mathcal{D}\mathcal{C}_{1,2}}=0.3$}
    & \REV{$\gamma^{\mathcal{C}_{2}\mathcal{D}}=0.003$} 
\end{tabular}
\end{table}

\subsection{Isolated road users}

\begin{figure}[t]
\centering
\includegraphics[scale=0.29]{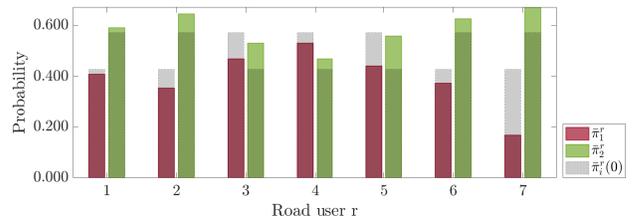}
\caption{\REV{Stationary state decision probabilities to \textit{Yield} (red, left column) and \textit{Go} (green, right column) of RUs 1-7 in the isolated case.}}
\label{fig:isolMarg}
\end{figure}

In summary, we describe two confident groups, $\mathcal{C}_{1}$ and $\mathcal{C}_{2}$, that interact with the antagonistic, uncertain group $\mathcal{D}$.
When examining how $\mathcal{D}$ acts depending on the attractive and repulsive forces, we consider the case of isolated RUs as a baseline. The state probabilities will follow the rate distributions in the individual rate matrices, which are biased toward the preferred states in Table \ref{tab:agentParams}.
\REV{The evolution of the marginalized decision probabilities can be found by solving (\ref{eq:margModelDE}) with $A_{m}$, $R_{m}$ and $E_{m}$ identically set to zero, although we use (\ref{eq:margModelSS1}) subject to (\ref{eq:margModelSS2}) to find their stationary values directly. In Fig. \ref{fig:isolMarg}, we compare the grey, arbitrarily selected initial state probabilities $\pi_{i}^{r}(0)$ to their stationary state values $\bar{\pi}_{i}^{r}$ for a clear visual comparison.} Notably, the highest state probability of each RU corresponds to its preferred state given in Table \ref{tab:agentParams}.
\subsection{Attractive and repulsive forces}
We now add the force functions and compare the probabilities in Fig. \ref{fig:isolMarg} to two cases. First, we add the attractive force to the marginalized model, so that $R_{m}$ and $E_{m}$ are set to zero in (\ref{eq:margModelDE}). Second, we include both forces, evaluating (\ref{eq:margModelDE}) in full. As before, we derive and compare the stationary states of (\ref{eq:margModelDE}), obtained through (\ref{eq:margModelSS1}) and (\ref{eq:margModelSS2}) in both cases.

\begin{figure}[b]
     \centering
     \subfloat[][Personal preferences with attractive force.]{\includegraphics[scale=0.29]{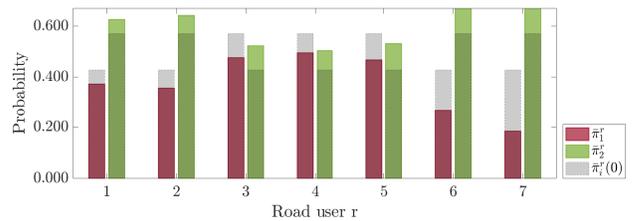}\label{fig:attMarg}}\\
     \subfloat[][Personal preferences with attractive and repulsive forces.]{\includegraphics[scale=0.29]{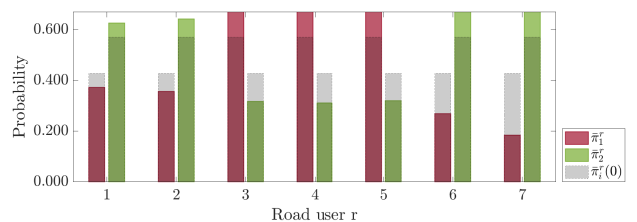}\label{fig:allMarg}}
     \caption{\REV{Stationary state probabilities to \textit{Yield} (red, left column) and \textit{Go} (green, right column) in two cases.}}
     \label{steady_state}
\end{figure}


Fig. \ref{fig:attMarg} describes the state probabilities in the first case.
\REV{Although each cyclist in $\mathcal{C}_{1}$ and $\mathcal{C}_{2}$ is already more likely to choose \textit{Go} in isolation, the probability of this decision increases even further for both cyclist $1$ and $6$. Moreover, driver $4$ in $\mathcal{D}$ may now select \textit{Go} or \textit{Yield} with equal probability, due to attractive influence. The behavior of $\mathcal{C}_{1}$ and $\mathcal{C}_{2}$ is explained by the fact that cyclists $1$ and $6$ are uncertain in relation to their group members, implying that one RU exerts a stronger attractive force on the other. In comparison, all drivers in $\mathcal{D}$ are equally uncertain, and the behavior of driver $4$ is instead an effect of peer pressure. The total attractive force is greater towards \textit{Go} than towards \textit{Yield}, and driver $4$ is thus slightly encouraged to go, from observing the other drivers.}


In the case of both attractive and repulsive forces, the decision state probabilities of the RUs are shown in Fig. \ref{fig:allMarg}. 
\REV{It can be seen that cyclists 1, 2, 6 and 7 still prefer \textit{Go}, as their state probability distributions are unchanged compared to the attractive force case in Fig. \ref{fig:attMarg}.
However, drivers 3, 4 and 5 now clearly prefer \textit{Yield}. This polarization is due to the almost uni-directional repulsive forces from $\mathcal{C}_{1}$ and $\mathcal{C}_{2}$ to $\mathcal{D}$, but also to the high average confidence of $\mathcal{C}_{1}$ and $\mathcal{C}_{2}$ and the high uncertainty of $\mathcal{D}$. The drivers in $\mathcal{D}$ respect the decisions of $\mathcal{C}_{1}$ and $\mathcal{C}_{2}$, and choose to yield. Thus, our model can be used to predict the most likely decisions of the road users in \textit{Problem 1} in Section \ref{sec:network}. For any initial probability distribution, the same model can also describe the evolution of the decision probabilities to their stationary values.}

Generally, our model predicts how uncertain agents may change their decisions in the presence of other agents. However, we would like to filter out unrealistic decisions that are too extreme compared to a decision made in isolation. Assume that the network model rate matrices $A_{0}$ and $R_{0}$ produce equally probable stationary states. As individual preferences are defined in $Q_{0}$, transition rates to network states that include unrealistic decisions can be lowered, implying that the corresponding diagonal elements in $Q_{0} + A_{0} + R_{0}$ are reduced. Diagonal elements reflect the propensity to leave each state \cite{cassandras2008introduction}, so the likelihood of leaving unrealistic network states can be increased due to personal preferences. This affects the stationary network probabilities, implying that unrealistic decisions can be visible also in its marginalization.

\section{Conclusions}
\label{sec:conclusions}
\REV{We have used Markovian opinion dynamics as a framework for modeling decision processes between stochastic agents, inspired by a traffic intersection problem in which road users collectively determine who should yield or go}. Agents are divided into groups, and emerging behaviors such as peer pressure and polarization regarding decisions are modeled using attractive and repulsive forces. We present a marginalization of our model, in the style of \cite{bolzern2018opinion} and \cite{bolzern2019opinion}, to reduce the state space dimension. With our formulation, collective decision processes with two types of interactions can be modeled using Markovian opinion dynamics.

\REV{In the intersection problem, our model expresses a high probability that cyclists will go through the intersection along with other, confident cyclists. Uncertain drivers are however predicted to yield, in order to avoid collisions.} 
\REV{In vehicle path planning problems, the stationary solution of our method could be used to predict road user behavior based on a complete traffic scene, whereas the transient solution may be used to express how such a prediction evolves over time.} 

\REV{Future work includes conducting a thorough comparison between our method and existing models for behavior prediction in traffic scenes,}
\REV{and investigating how traffic data can be used to estimate influence parameters. Specifically, a topic of our ongoing research is predicting the probabilities that road users in the \textit{InD} intersection dataset \cite{InDdataset} yield or go.}




\addtolength{\textheight}{-12cm}   




\bibliographystyle{IEEEtran}
\bibliography{IEEEabrv,references.bib}

\begin{thebibliography}{10}
\providecommand{\url}[1]{#1}
\csname url@rmstyle\endcsname
\providecommand{\newblock}{\relax}
\providecommand{\bibinfo}[2]{#2}
\providecommand\BIBentrySTDinterwordspacing{\spaceskip=0pt\relax}
\providecommand\BIBentryALTinterwordstretchfactor{4}
\providecommand\BIBentryALTinterwordspacing{\spaceskip=\fontdimen2\font plus
\BIBentryALTinterwordstretchfactor\fontdimen3\font minus
  \fontdimen4\font\relax}
\providecommand\BIBforeignlanguage[2]{{%
\expandafter\ifx\csname l@#1\endcsname\relax
\typeout{** WARNING: IEEEtran.bst: No hyphenation pattern has been}%
\typeout{** loaded for the language `#1'. Using the pattern for}%
\typeout{** the default language instead.}%
\else
\language=\csname l@#1\endcsname
\fi
#2}}

\bibitem{batkovic2021robust}
I.~Batkovic, U.~Rosolia, M.~Zanon, and P.~Falcone, ``A robust scenario mpc
  approach for uncertain multi-modal obstacles,'' \emph{IEEE Control Systems
  Letters}, vol.~5, no.~3, pp. 947--952, 2021.

\bibitem{noorazar2020survey}
H.~Noorazar, K.~R. Vixie, A.~Talebanpour, and Y.~Hu,
  ``\BIBforeignlanguage{English}{From classical to modern opinion dynamics},''
  \emph{\BIBforeignlanguage{English}{International Journal of Modern Physics
  C}}, vol.~31, no.~7, 2020.

\bibitem{degroot1974consensus}
M.~H. Degroot, ``\BIBforeignlanguage{English}{Reaching a consensus},''
  \emph{\BIBforeignlanguage{English}{Journal of the American Statistical
  Association}}, vol.~69, no. 345, pp. 118--121, 1974.

\bibitem{friedkin1990social}
N.~E. Friedkin and E.~C. Johnsen, ``\BIBforeignlanguage{English}{Social
  influence and opinions},'' \emph{\BIBforeignlanguage{English}{The Journal of
  Mathematical Sociology}}, vol.~15, no. 3-4, pp. 193--206, 1990.

\bibitem{deffuant2001mixing}
G.~Deffuant, D.~Neau, F.~Amblard, and G.~Weisbuch, ``Mixing beliefs among
  interacting agents,'' \emph{Advances in Complex Systems}, no.~3, p.~11, 2001.

\bibitem{asavathiratham2001influence}
C.~Asavathiratham, S.~Roy, B.~Lesieutre, and G.~Verghese,
  ``\BIBforeignlanguage{English}{The influence model},''
  \emph{\BIBforeignlanguage{English}{IEEE Control Systems Magazine}}, vol.~21,
  no.~6, pp. 52--64, 2001.

\bibitem{vanmieghem2011sis}
P.~Van~Mieghem, ``The n-intertwined sis epidemic network model.''
  \emph{Computing}, vol.~93, no. 2-4, pp. 147 -- 169, 2011.

\bibitem{banisch2012opinion}
S.~Banisch, R.~Lima, and T.~Araújo, ``\BIBforeignlanguage{English}{Agent based
  models and opinion dynamics as markov chains},''
  \emph{\BIBforeignlanguage{English}{Social Networks}}, vol.~34, no.~4, pp.
  549--561, 2012.

\bibitem{bolzern2014consensus}
P.~Bolzern, D.~Cerotti, P.~Colaneri, and M.~Gribaudo, ``Probabilistic consensus
  in markovian multi-agent networks,'' in \emph{2014 European Control
  Conference (ECC)}, 2014, pp. 558--563.

\bibitem{bolzern2018opinion}
P.~Bolzern, P.~Colaneri, and G.~{De Nicolao}, ``Opinion dynamics in social
  networks with heterogeneous markovian agents,'' in \emph{2018 {IEEE}
  Conference on Decision and Control (CDC)}, Miami, USA, Dec. 2018, pp.
  6180--6185.

\bibitem{bolzern2019opinion}
P.~{Bolzern}, P.~{Colaneri}, and G.~{De Nicolao}, ``Opinion influence and
  evolution in social networks: A markovian agents model,'' \emph{Automatica},
  vol. 100, pp. 219--230, 2019.

\bibitem{bolzern2020opinion}
P.~Bolzern, P.~Colaneri, and G.~De~Nicolao, ``Opinion dynamics in social
  networks: The effect of centralized interaction tuning on emerging
  behaviors,'' \emph{IEEE Transactions on Computational Social Systems},
  vol.~7, no.~2, pp. 362--372, 2020.

\bibitem{bolzern2021effect}
P.~Bolzern, P.~Colaneri, and G.~{De Nicolao}, ``Effect of social influence on a
  two party election: A markovian multi-agent model,'' pp. 1--1, 2021.

\bibitem{farooqi2019railway}
H.~Farooqi, G.~P. Incremona, and P.~Colaneri,
  ``\BIBforeignlanguage{English}{Railway collaborative ecodrive via dissension
  based switching nonlinear model predictive control},''
  \emph{\BIBforeignlanguage{English}{European Journal of Control}}, vol.~50,
  pp. 153--160, 2019.

\bibitem{noorazar2018energy}
H.~Noorazar, M.~J. Sottile, and K.~R. Vixie, ``\BIBforeignlanguage{English}{An
  energy-based interaction model for population opinion dynamics with topic
  coupling},'' \emph{\BIBforeignlanguage{English}{International Journal of
  Modern Physics C}}, vol.~29, no.~11, 2018.

\bibitem{antal2006social}
T.~Antal, P.~L. Krapivsky, and S.~Redner, ``\BIBforeignlanguage{English}{Social
  balance on networks: The dynamics of friendship and enmity},''
  \emph{\BIBforeignlanguage{English}{Physica D: Nonlinear Phenomena}}, vol.
  224, no. 1-2, pp. 130--136, 2006.

\bibitem{altafini2012dynamics}
C.~Altafini, ``Dynamics of opinion forming in structurally balanced social
  networks,'' \emph{PLOS ONE}, vol.~7, no.~6, pp. 1--9, 06 2012.

\bibitem{yang2019opinion}
Y.~Yang and Y.~Song, ``\BIBforeignlanguage{English}{A novel interpretation for
  opinion consensus in social networks with antagonisms},''
  \emph{\BIBforeignlanguage{English}{IEEE Access}}, vol.~7, pp.
  51\,475--51\,483, 2019.

\bibitem{cassandras2008introduction}
C.~G. Cassandras and S.~Lafortune, ``Markov chains,'' in \emph{Introduction to
  discrete event systems}, 2nd~ed.\hskip 1em plus 0.5em minus 0.4em\relax
  Springer, 2008, ch.~7, pp. 368--428.

\bibitem{InDdataset}
J.~Bock, R.~Krajewski, T.~Moers, S.~Runde, L.~Vater, and L.~Eckstein, ``{The
  inD Dataset: A Drone Dataset of Naturalistic Road User Trajectories at German
  Intersections},'' in \emph{2020 IEEE Intelligent Vehicles Symposium (IV)},
  2020, pp. 1929--1934.

\end{thebibliography}

\end{document}